\documentclass[reprint,aps,pre,amsmath,amssymb,superscriptaddress,showpacs]{revtex4-1}
\usepackage{graphicx}

\begin{document}

\title{Growth principles of natural hypergraphs}

\author{Alexei Vazquez}
\email{alexei@nodeslinks.com}
\affiliation{Nodes \& Links Ltd, Salisbury House, Station Road, Cambridge, CB1 2LA, UK}


\begin{abstract}
Several systems can be represented by hypergraphs, an extension of graphs with associations between any number of vertices. These natural hypergraphs doe not appear at once. They are generated by some dynamical process of hypergraph evolution. Here I investigate what are the minimal growth principles of natural hypergraphs. I postulate edge duplication and vertex addition at edge duplications as the key principles of hypergraph growth. The implementation of these two principles induce the emergence of preferential attachment, power law degree distribution, the small-world property, high clustering coefficient and the founder effect. This work clarifies the distinction between principles, emergent properties and context specific details in the context of hypergraph growth dynamics.
\end{abstract}

\maketitle

\section{Introduction}

In recent years there has been an increased interest in higher order associations beyond pairwise interactions \cite{battiston20,battiston21}.  Higher order associations have been represented by hypergraphs and simplicial complexes. Hypergraphs are combinatorial objects that generalizes graphs to include associations between more than two elements. Simplicial complexes are families of higher order associations that are closed under taking subsets (e.g., nodes, edges, triangles, \ldots) and they are the building blocks of topological spaces. These mathematical constructions have been around for quite some time. The novelty is in how they appear in natural and human related systems \cite{wang10, petri18, iacopini19, coutinho20, civilini21, stonge21, neuhauser21}, how they can be generated by random ensembles \cite{chodrow21, young21} and how they impact the behavior of process on such structures \cite{vazquez09hg, ghoshal09, courtney18}.

When studying models of natural or human related systems we should make a distinction between principles, emergent properties and details. Let us take as an example preferential attachment in the context of networks \cite{barabasi99}. There is no doubt preferential attachment is a distinctive feature in the dynamics of real networks. Yet, preferential attachment is an emergent property of local evolution rules on networks \cite{vazquez03local}. When navigating a network the transversed links bias the walk to nodes with a higher number of neighbors (degree), with a bias proportional to the degree. New links generated by local walks on an existing network lead to preferential attachment \cite{vazquez01rs}. Making copies of existing nodes (e.g., web pages or proteins) generates preferential attachment as well \cite{kleinberg99,pastor-satorras03,vazquez03dup,chung03}. The local rules do more, they generate the small world property, high clustering coefficient and degree correlations \cite{vazquez03local}.

In line with the timeline of growth models of networks, preferential attachment has been extended to the context hypergraph growth models \cite{wang10}. Yet again, as in the context of networks, I would expect that preferential attachment is an emergent property of dynamical systems with a hypergraph structure. The basic mechanism should be rooted on some local dynamics.

Here I investigate the principles of natural hypergraphs growth dynamics. I have gotten inspiration from previous work on the topic \cite{courtney18,kovalenko21}. My contribution is to establish a clear separation between principles and consequence, cause and effect. I start in Section \ref{principles} postulating the principles of hypergraph evolution. I follow in Section \ref{models} with an analysis of hypergraph properties derived from those principles. The emphasis is on the emergent properties: preferential attachment \cite{barabasi99}, scale-free degree distribution \cite{albert99}, small-world and high clustering coefficient \cite{watts98}. In Section \ref{founder} I elaborate on the founder effect as another emergent property. Finishing with concluding remarks in Section \ref{conclusions}.

\section{Principles}
\label{principles}

Preferential attachment and scale-free degree distributions are consequencse of network dynamics with local evolution rules \cite{vazquez03local}.  In a nutshell, whenever we transverse an edge we are performing a bias sampling towards vertices with higher degrees. Selecting an edge, rather than a vertex, is what drives the preferential attachment to vertices with higher degree. Since many of the networks studied in the past are actually projections of latent hypergraphs (e.g., co-autorships and metabolic networks), I expect that similar principles apply to the dynamics of natural and human related hypergraphs.

A hypergraph $H(V,E)$ is composed by a set of vertices $V$ and a set of edges $E$, where edges are sets with one or more vertices. When there are not two edges with the same vertex composition we call it a simple hypergraph, or a multi-hypergraph otherwise. Following the principles of network evolution, I focus on local and edge centric evolution principles of hypegraphs. Local means that the evolution rules are based on one or a few edges in a local neighborhood of the hypergraph. The simplest local rule of hypergraph growth is edge duplication
\begin{equation}
e \rightarrow 2e\  .
\label{dup}
\end{equation}
There are several natural and human related scenarios that are represented in this rule.  Starting with the obvious resemblance with asexual reproduction by cell division. If vertices are genes and edges are cells then Eq. (\ref{dup}) is a model of cell division. In the context of human behavior an edge can be an instance of a recurrent activity involving a group of people (the vertices). Examples include scientific publications, scientific meetings and working days at institutions departments. Notice that Eq. (\ref{dup}) generates multi-hypergraphs or temporal hypergraphs if we add a time stamp label to each edge.

We need a second rule to model the addition of new vertices. The simplest of local edge centric rules is duplication with vertex addition at the duplicate
\begin{equation}
e \rightarrow e + e\cup\{v\}\ .
\label{dupadd}
\end{equation}
In asexual reproduction this rule represents gene duplication, an event that can happen during the genome replication preceding cell division. In the context of human behavior Eq. (\ref{dupadd}) represents a new individual joining a group at their next group meeting. We could spell further evolution rules involving two or more edges that are adjacent (thus local) leading to edge composition mixing. Here I will stop at the two simplest rules (\ref{dup})-(\ref{dupadd}) and investigate the structure of hypergraphs they generate.

\begin{figure}[t]
\includegraphics[width=3.3in]{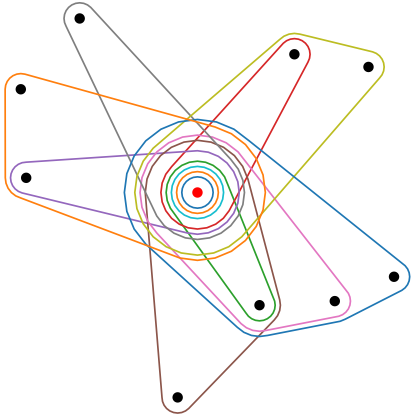}
\caption{Hypergraph instance generated with the implementation of rules (\ref{dup})-(\ref{dupadd}) (App. \ref{app_dupadd}), using as input $N=10$ vertices, vertex growth rate $\mu=0.5$ and a starting hypergraph consisting of one edge with one vertex. Visualized with Python package HyperNetX.}
\label{fig_ex1}
\end{figure}

\section{Emergent properties}
\label{models}

To investigate the emergent properties from the hypergraph growth principles I introduce a specific model. At time $t=0$ we start with hypergraph $H_0$. At each subsequent time step, we select an edge $e$ sampled with equal probability among all current edges. To this edge we apply (\ref{dup}) with probability $1-\mu$ or (\ref{dupadd}) otherwise. When (\ref{dup}) is applied, we create a new edge $e^\prime = e$ with the exact same vertex composition. When (\ref{dupadd}) is applied, we add a new vertex $n\rightarrow n+1$ and a new edge $e^\prime = e\cup\{v(n+1)\}$ with the composition of $e$ plus the new vertex. We continue until the hypergraph has $n=N$ vertices. Since new vertices are added with probability $\mu$, $\mu$ represents the vertex growth rate per edge duplication. Each hypergraph ensemble is parametrized by $\{H_0,\ \mu,\ N\}$. 

A Python implementation is listed in Appendix \ref{app_dupadd}. A hypergraph example generated with this code is shown in Fig. \ref{fig_ex1}. The star-like shape already anticipates the heterogeneity in the vertices connectivity and the small world property.

\subsection{Preferential attachment}

The degree of a vertex is the number of edges containing that vertex and it will be denoted by $k$. At each step one edge is selected for duplication, with or without vertex addition. The probability that a vertex with degree $k$ is in the selected edge, and therefore increases its degree, is
\begin{equation}
\Pi_k(t) = \frac{k}{E(t)}
\label{Pik}
\end{equation}
where $E(t)$ is the number of edges at step $t$. That is preferential attachment: new edges are added at vertices with a bias by their degree. 

\begin{figure}[t]
\includegraphics[width=3.3in]{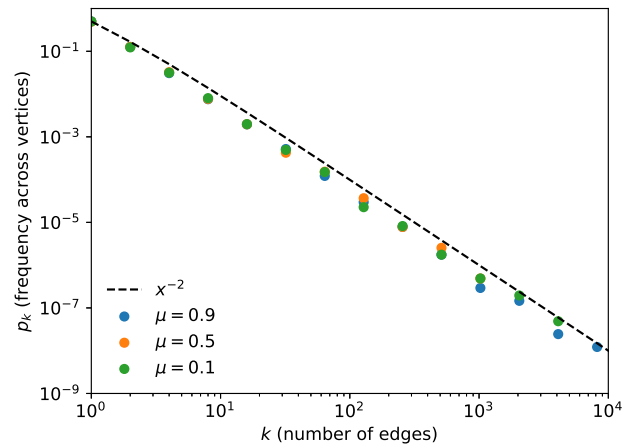}
\caption{Degree distribution of hypergraph instances generated with the implementation of rules (\ref{dup})-(\ref{dupadd}) (App. \ref{app_dupadd}), using as input $N=10000$ vertices, the vertex growth rate $\mu$ indicated in the legend and a starting hypergraph consisting of one edge with one vertex. The dashed line is the expected degree distribution $p_k=1/(k(k+1))$.}
\label{fig_pk}
\end{figure}

\begin{figure}[t]
\includegraphics[width=3.3in]{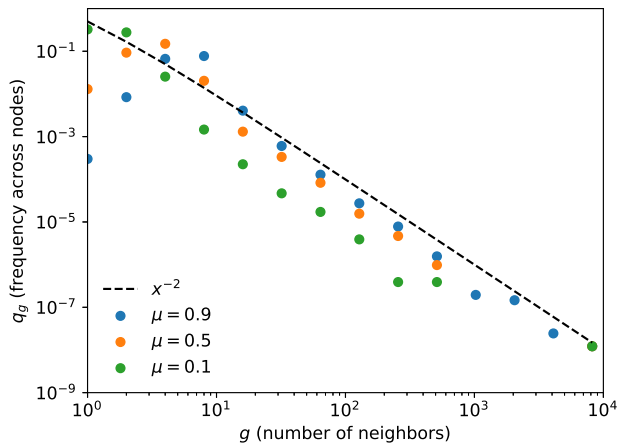}
\caption{Number of nearest neighbors distribution of hypergraph instances generated with the implementation of rules (\ref{dup})-(\ref{dupadd}) (App. \ref{app_dupadd}), using as input $N=10000$ vertices, the vertex growth rate $\mu$ indicated in the legend and a starting hypergraph consisting of one edge with one vertex. The dashed line is the expected degree distribution $p_k=1/(k(k+1))$.}
\label{fig_pg}
\end{figure}

\subsection{Power law degree distribution}

For large hypergraphs ($t\gg1$), the rate equation describing the evolution of the number $n_k(t)$ of vertices with degree $k$ is
\begin{eqnarray}
n_k(t+1) &=& n_k(t) + \Pi_{k-1}(t) n_{k-1}(t) -  \Pi_k(t) n_{k}(t)\nonumber\\
&& + \mu \delta_{k1}\ .
\label{nkt}
\end{eqnarray}
Since an edge is added at each step and vertices are added at rate $\mu$, $E(t)\approx t$ and $N(t)\approx \mu t$ for $t\gg1$. In such a case Eq. (\ref{nkt}) has the steady state solution $n_k(t) = \mu t p_k$, where $p_k$ satisfies the recursive equation
\begin{equation}
p_k = (k-1) p_{k-1} -  k p_k + \delta_{k1}
\label{pkpk}
\end{equation}
Iterating this equation with the boundary condition $p_{-1}=0$ I obtain
\begin{equation}
p_k = \frac{1}{k(k+1)} 
\label{pk}
\end{equation}
The degree distribution of hypergraphs generated by (\ref{dup})-(\ref{dupadd}) have a power low tail with exponent -2. This is corroborated by numerical simulations (Fig. \ref{fig_pk}).

In many applications the hypergraph is projected into a graph, where two vertices are connected if there is at least one edge containing them (on-vertex projection). The vertex degree in this graph is the number of nearest neighbors, the number of other vertices that can be reached via the edges containing the vertex.  Although I do not have an analytical derivation for the number of neighbors distribution, I expect it is characterized by the same power law tail with exponent -2. This seems to be the cased based on numerical simulations (Fig. \ref{fig_pg})

\begin{figure}[t]
\includegraphics[width=3.3in]{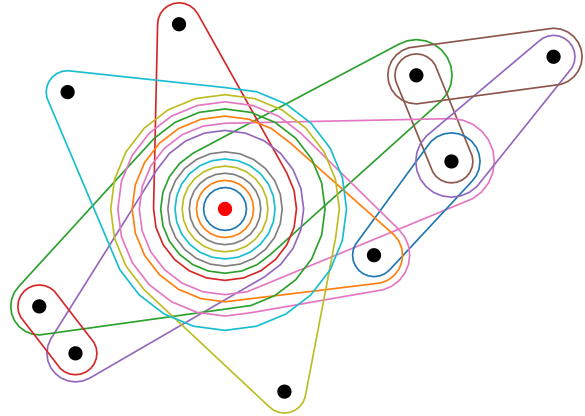}
\caption{Hypergraph instance generated with the implementation of rules (\ref{dup})+(\ref{dupaddbreak}) (App. \ref{app_dupaddbreak}) with maximum edge size $s_c=2$, using as input $N=10$ vertices, vertex growth rate $\mu=0.5$ and starting from a hypergraph consisting of one edge with one vertex.}
\label{fig_ex2}
\end{figure}

\begin{figure}[t]
\includegraphics[width=3.3in]{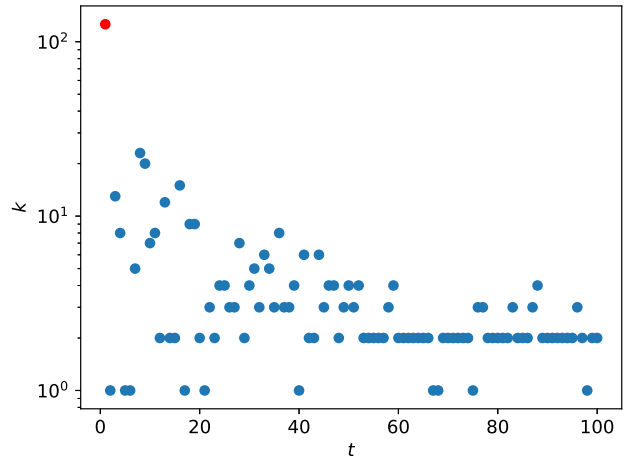}
\caption{Scatter plot of the vertices degree as a function of the time stamp when they were added, for a hypergraph 
instance generated with the implementation of rules (\ref{dup})+(\ref{dupaddbreak}) (App. \ref{app_dupaddbreak}). The input parameters were $N=100$ vertices, vertex growth rate $\mu=0.5$, maximum edge size $s_c=2$ and starting hypergraph consisting of one edge with one vertex. The red symbol is the founder.}
\label{fig_founder}
\end{figure}

\subsection{Founder effect}
\label{founder}

Visual inspection of Fig. \ref{fig_ex1} indicates that the seeding vertex is part of all edges. This is indeed the case. If we start with one edge containing one or more vertices then all other edges will contain those starting vertices. The founders persist in all edges.

This behavior can be relaxed if we specify further rules. For example, during the growth of organizations large departments are split and new managers are brought in to maintain communication between the new departments. This can be modeled by duplication with edge split and vertex addition
\begin{equation}
e \rightarrow \left\{
\begin{array}{ll}
e + e\cup\{v\} & {\rm if}\ |e|<e_c\ ,\\ 
e + e_1\cup\{v\} + e_2\cup\{v\} & {\rm otherwise}\ ,
\end{array}
\right.
\label{dupaddbreak}
\end{equation}
where $e_1 + e_2 = e$.
A specific implementation with random and balance edge split is found in Appendix \ref{app_dupaddbreak}. Figure \ref{fig_ex2} shows an example with edges constrained to size 2. The vertex in the far right does not share an edge with the founder (in red). Yet, the founder is still connected to most vertices.

A founder is basically a member that joined the system early in its evolution and it has a high connectivity at later times. This is illustrated in Fig. \ref{fig_founder}, with the founder highlighted in red. In fact, that plot is an operational method to identify putative founders. They are found in the upper left corner.

\begin{figure}[t]
\includegraphics[width=3.3in]{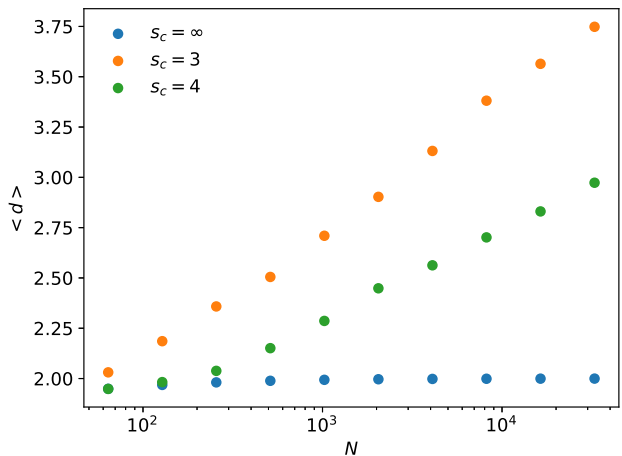}
\caption{Semi-logarithmic scaling of the average shortest path distance between vertices with the number of vertices. The hypergraphs instances were generated generated with the implementation of rules (\ref{dup})+(\ref{dupaddbreak}) (App. \ref{app_dupadd}), using as input $N=10000$ vertices, vertex growth rate $\mu=0.5$, the maximum edge sizes $s_c$ indicated in the legend and a starting hypergraph consisting of one edge with one vertex.}
\label{fig_dn}
\end{figure}

\subsection{Small-world property}

The founder effect induces the small-world property. The hypergraphs generated by (\ref{dup})-(\ref{dupadd}) have diameter 2. Any two vertices are at most two edges away, each edge containing the founder. This singular behavior of constant diameter is relaxed after adding the edge split rule (\ref{dupaddbreak}). When a cap is imposed in the edge size there is a logarithmic increase in the average shortest path distance with increasing the number of vertices (Fig. \ref{fig_dn}), the typical behavior of small-world networks \cite{amaral00}. Furthermore, the larger the maximum allowed edge size the smaller the hypergraph, getting closer to the unconstrained case (Fig. \ref{fig_dn}, blue symbols).

\subsection{High clustering coefficient}

The hypergraph extension to include edges with size larger than 2 induces a high clustering coefficient in the associated graph on-vertex projection. Not surprisingly hypergraphs evolving through the growth principles introduced here  have a high clustering coefficient. The numerical estimates of the average clustering coefficient are above 0.5 for the range of parameters investigated (see Table \ref{tab_r}).

\begin{table}

\begin{tabular}{l|l|l|l|l|r}
$H_0$ &  $s_c$ &  $\mu$ &  $\langle d\rangle$ &  $\langle c\rangle$ &   $r$ \\
\hline
   H1 &      2 &    0.1 &                2.25 &                0.55 & -0.53 \\
   H1 &      2 &    0.5 &                4.54 &                0.76 & -0.07 \\
   H1 &      2 &    0.9 &                4.69 &                0.74 & -0.07 \\
   H1 &      0 &    0.1 &                2.00 &                0.55 & -0.59 \\
   H1 &      0 &    0.5 &                1.99 &                0.91 & -0.22 \\
   H1 &      0 &    0.9 &                1.99 &                0.92 & -0.15 \\
  H10 &      2 &    0.1 &                2.14 &                0.93 &  0.07 \\
  H10 &      2 &    0.5 &                3.31 &                0.80 &  0.05 \\
  H10 &      2 &    0.9 &                4.25 &                0.75 &  0.18 \\
  H10 &      0 &    0.1 &                1.98 &                0.98 & -0.01 \\
  H10 &      0 &    0.5 &                1.97 &                0.97 & -0.03 \\
  H10 &      0 &    0.9 &                1.97 &                0.95 & -0.03 \\
\end{tabular}

\caption{Properties of hypergraph instances generated with the implementation of rules (\ref{dup})+(\ref{dupaddbreak}) (App. \ref{app_dupaddbreak}), using as input $N=10000$ vertices, different vertex growth rates $\mu$, different maximum edge sizes $s_c$ and starting hypergraph $H1$ or $H10$. H1 is a hypergraph with an edge containing one vertex. H10 is is a hypergraph with an edge containing 10 vertices.}
\label{tab_r}
\end{table}

\subsection{Degree correlations}

Degree correlations inform about degree mixing between connected vertices \cite{pastor-satorras01,newman02}. Degree correlations are quantified by the Pearson correlation of the degree of connected vertices, denoted by $r$ \cite{newman02}. When $r>0$ we say there is assortative mixing, whereby connected vertices tend to have similar degrees. When $r<0$ we say there is disassortative mixing, whereby connected vertices tend to have dissimilar degrees. For hypergraphs we can use the same quantity, interpreting connected vertices as vertices sharing at least one hyper-edge.

The hypergraphs generated by (\ref{dup})-(\ref{dupadd}) or (\ref{dup})+(\ref{dupaddbreak}) have variable values of $r$ ranging from negative to positive values (Tab. \ref{tab_r}). $r$ is modulated by the founder effect. For example, when the starting hypergraph has one edge containing one vertex then $r<0$ for all values of $\mu$ and the maximum edge size $s_c$ that I have simulated.  In contrast, when the starting hypergraph has one edge containing 10 vertices then $r>0$ for $s_c=2$ but $r<0$ for $s_c=\infty$, for all simulated values of $\mu$. Therefore degree correlations are sensitive to parameters and initial conditions. They are context depending.

\section{Conclusions}
\label{conclusions}

Edge duplication and vertex addition at edge duplication are natural principles of hypergraph evolution. The universal properties of real world networks that are projections of hypergraphs are emergent properties from these two growth principles.

The evolution of natural or human related hypergraphs is richer. I have illustrated the impact of adding a constraint on the maximum edge size. Higher order evolution rules can involve two or more edges.  These variations will modulate the emergent properties described above. They could modify the exponent of the power law tail in the degree distribution or increase the diameter. Yet, as long as the hypergraph growth principles of edge duplication and vertex addition at edge duplication are part of the system dynamics the emergent properties will persist.

The founder effect modulates the hypergraph properties. The founder effect is a typical property of natural systems that grow from a small initial seed. Biological populations of organisms is the canonical example. The founder effect is observed during the creation of new academic fields or institutions. Based on this work, the founder effect is an emergent property of the hypergraph growth principles.

\section*{Acknowledgements}

Nodes \& Links did not have any role in the study design, data collection and analysis, decision to publish, or preparation of the manuscript.

\appendix

\section{}
\label{app_dupadd}

Implementation of (\ref{dup})-(\ref{dupadd}) in Python
\begin{verbatim}
import numpy as np
import hypernetx as hnx
def HGP(n, mu, E0, seed):
    # hypergraph by growth principles
    np.random.seed(seed)
    E = E0
    i = max([j for e in E for j in e]) + 1
    while i<n:
        e = E[np.random.randint(len(E))]
        if np.random.random() < mu:
            E.append(e + [i])
            i += 1
        else:        
            E.append(e)
    return E
E = HP1(10, 0.5, [[0]], 100)
H = hnx.Hypergraph(E)
hnx.drawing.draw(H)
\end{verbatim}

\section{}
\label{app_dupaddbreak}

Implementation of (\ref{dup})+(\ref{dupaddbreak}) inPpython
\begin{verbatim} 
import numpy as np
import hypernetx as hnx
def HGPSC(n, mu, E0, seed, sc):
    # hypergraph growth principles
    #   with edge size constraint
    np.random.seed(seed)
    E = E0
    i = max([j for e in E for j in e]) + 1
    while i<n:
        e = E[np.random.randint(len(E))]
        if np.random.random() < mu:
            if len(e)<sc:
                E.append(e + [i])
            else:
                split = np.array_split(
                    np.random.permutation(e),2)
                E.append(list(split[0])+[i])
                E.append(list(split[1])+[i])            
            i += 1
        else:        
            E.append(e)
    return E
E = HGPSC(10, 0.5, [[0]], 100, 2)
H = hnx.Hypergraph(E)
hnx.drawing.draw(H)
\end{verbatim}

\bibliographystyle{apsrev4-1}


\bibliography{network.bbl}

\end{document}